\documentclass[12pt]{report}
\usepackage[utf8]{inputenc}
\usepackage[english]{babel}
\usepackage{gensymb}
\usepackage{amssymb}
\usepackage{graphicx}
\usepackage{psfrag}
\usepackage{afterpage}
\usepackage[rflt]{floatflt}

\begin{document}

\centerline{\Large \bf How to deal with PCR composition problem at $E_0 \gtrsim 10^{17}$ eV}
\vspace{1.5cm}

\centerline{\it Galkin V.I., Anokhina A.M., Bakhromzod R., Mukumov A.}
\vspace{0.5cm}

\centerline{\it Faculty of Physics of Lomonosov Moscow State University}
\centerline{\it  Leninskie Gory, Moscow, 119991,Russian Federation}
\centerline{\it  E-mail: v\_i\_galkin@mail.ru, anokhanna@rambler.ru}
\vspace{1.5cm}

\centerline{\bf Abstract}
\vspace{0.5cm}

Basic ideas of muon tracker technique for the solution of primary cosmic ray (PCR)
composition problem in the energy range $10^{17}-10^{18}$ eV are presented. The approach uses MC simulation data made with CORSIKA6.990 for "Pamir-XXI" \hspace{0.1cm} site
conditions. Similar technology can certainly be developed for other observation levels and interaction models. One can probably extend it to much higher primary energies.
\vspace{1.5cm}

\centerline{\bf Introduction}
\vspace{0.5cm}

It is useless to repeat all the words that show the importance of the study of primary mass composition at super high energies covered by the extensive air shower (EAS) method. Many different groups using different detector set-ups
% [...]
 have dealt with the problem but the answer is still vague. The reason for such a poor result, from our viewpoint, lies in the inadequacy of experimental methods used. To be more precise, EAS characteristics that are experimentally measured ($X_{max}$, $N_{\mu}/N_{charged}$) do not contain enough information on the mass of the primary particle. A natural way out of such a situation is to look for more informative measures of the primary mass.

In 1-100 PeV energy range the EAS Cherenkov light angular distribution characteristics prove to be appropriate for the solution of mass composition problem [1,2]. Generally speaking, this technique can be used at higher energies as well but the detector array should be made much larger while still being rather dense and, thus, must be very expensive.

Here we present another approach for $E_0 \gtrsim 100$ PeV which seems to be more cost-effective and, probably, will work even for the extremely high energy showers.
\vspace{1.5cm}

\centerline{\bf Main idea of the new method}
\vspace{0.5cm}

EAS is a particle cascade originated by a primary cosmic ray, probably proton or other nucleus of super high energy. The primary particle undergoes a few strong interactions with air nuclei (N, O, Ar) giving life to quite a number of secondary particles. The most numerous EAS component is the electromagnetic one ($\gamma , e^+, e^-$) which is characterized by a {\it radiation length} (37 g/cm${}^2$) that is about 3 times smaller
%less
 than hadron interaction length. On the contrary, the range of the next to most numerous component, which is the muon one, can be much larger and for $E_{\mu} \sim$ 10 GeV is comparable with the shower length (10-30 km). One can conclude that the EAS muons with $E_{\mu} \ge$ 1 GeV can yield important information on the longitudinal shower development. In this aspect muons resemble the EAS Cherenkov light which proved to be the most informative component of all. %[...].
  Crucial question is how to collect the data on the PCR mass composition borne by muons, i.e. what characteristic of the muon component is sensitive to the primary mass and, thus, to be measured in EAS experiment.
 
Usually, the muon content $N_{\mu}$ of the shower is used as a measure of the primary particle mass. Unfortunately, any procedure for $N_{\mu}$ estimation involves the measurement of muon densities all over the area swept by EAS. The measurement is made by local detectors of limited area $\sim$1 m${}^2$ spread widely which results in rather large uncertainties of statistical (due to small detector area) and physical (due to cascade fluctuations) origin. Finally, it turns out that $N_{\mu}$ uncertainty kills its sensitivity to the primary mass. As the ultimate results of KASKADE-Grande [3] show, $N_{\mu}$ parameter is only capable of dividing all the events into two groups in primary mass which is definitely not enough to solve the composition problem.

Certainly, one should think of finding more sensitive parameters, preferably directly measurable ones. One of the possibilities is to use the similarity between the muons and Cherenkov light.

An optical detector observing an EAS from large ($\gtrsim$300 m) core distance reveals a Cherenkov light pulse presenting roughly an electron cascade curve. The latter can also be seen by an imaging telescope which field of view is divided into sensitive areas of small size
% $<1^{\circ}$
$\, <1 \degree$ in diameter (pixels).

Because relativistic muons can reach the observation level even from as high as the first interaction of the primary particle, one can think of some muon analogue of the imaging Cherenkov telescope, namely, of a muon tracker as a detector of muon angular distribution which should be sensitive to the primary mass. A priory it is not clear \\
a) whether such a sensor can distinguish even between primary proton and iron nucleus events, cascade fluctuations taken into account; \\
b) what the design of such a tracker should be; \\
c) how must the experimental data on muon angular distribution be processed in order the primary mass information to be extracted.

Full MC simulation presumably can answer the questions.
\vspace{1.5cm}

\centerline{\bf Artificial shower generation}
\vspace{0.5cm}

The present study is made in the framework of the "Pamir-XXI" project which determines the observation level (4250 m above sea level) and the primary energy range of interest ($10^{16}-10^{18}$ eV for secondary charged particle methods). "Pamir-XXI" will include an optical part [2] incorporating a net of fast detectors and imaging telescopes that, working together, are capable of dealing with PCR  mass composition problem in 100 TeV -- 100 PeV energy range. At higher energies the event rate with the basic (minimal) optical set-up will be low and one should think of a charged particle detector set-up to take over the task.

In view of all written above, one should first check $10^{17}-10^{18}$ eV energy range for a muon angular distribution sensitivity to the primary mass.

The following EAS samples were generated: \\
- 100 PeV proton (p), nitrogen (N) and iron nucleus (Fe) initiated vertical showers, sample volumes 50; \\
- 1 EeV p, N and Fe initiated vertical showers, sample volumes 30.

For the generation of all samples CORSIKA6.990/QGSJET01 [4] was used without thinning with energy threshold 1 GeV for all secondaries.

Also 100 PeV p, N and Fe initiated vertical shower samples of volume 30 were generated by CORSIKA6.990/QGSJET-II to analyze the interaction model effect on muon angular distribution.
\vspace{0.3cm}

\begin{figure}[h]
\includegraphics[width=44pc]{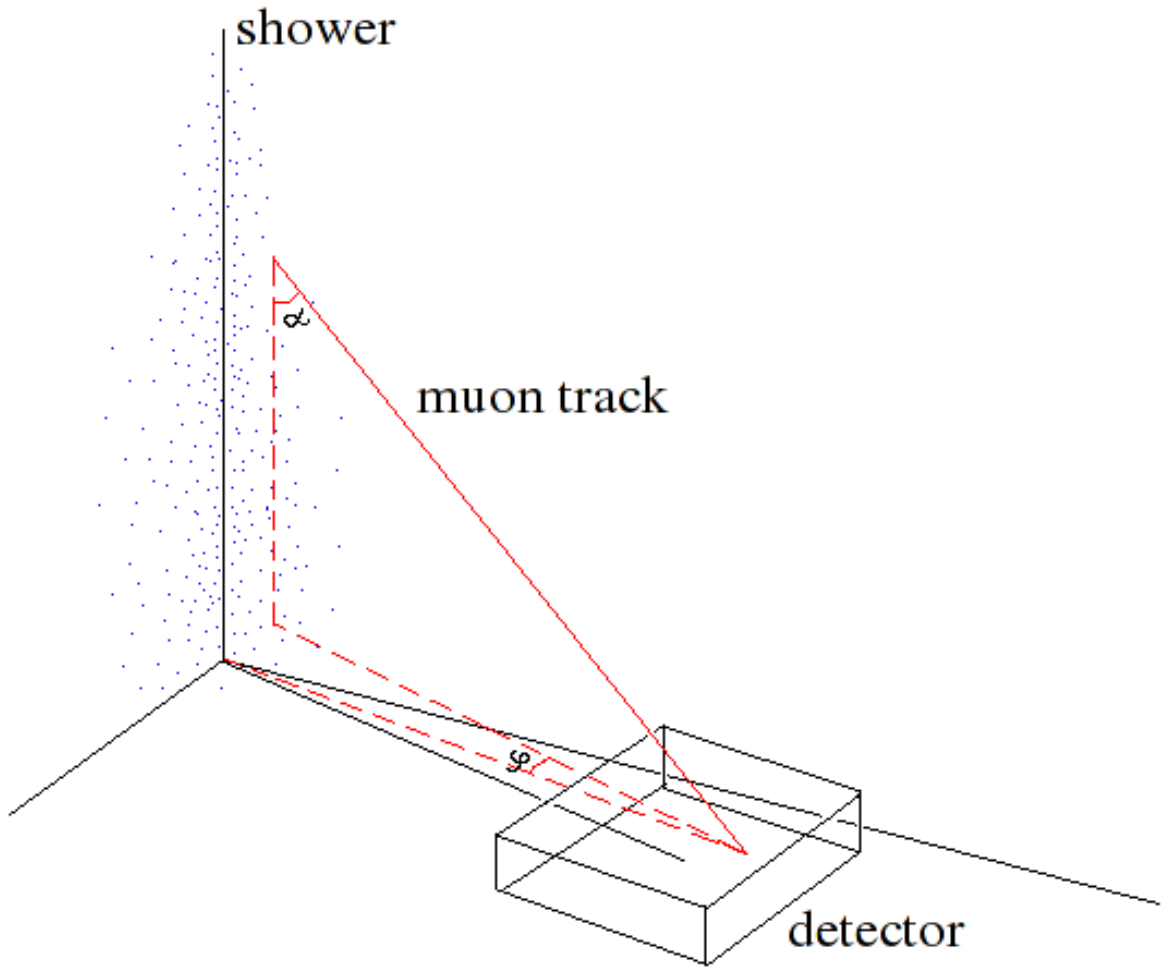}\hspace{2pc}%
\caption{\label{l1} Detection geometry for vertical shower: polar $\alpha$ and azimuthal $\phi$ of a muon track.}
\end{figure}

\vspace{1.0cm}
\centerline{\bf Muon angular distribution processing: how to}
\centerline{\bf distinguish between different primary nuclei}
\vspace{0.5cm}

The following considerations defined the choice of the main parameters of the muon tracker and data handling procedure (area, angular resolution, typical core distance, threshold energy, azimuthal angle limitations): \\
- tracker area should be large enough to detect a number of muons in each shower such that a histogram in muon polar angle $\alpha$ (with respect to the shower axis, see Fig.1) can be formed with bin contents fluctuating not too much; \\

\begin{figure}[h]
\includegraphics[scale=0.75]{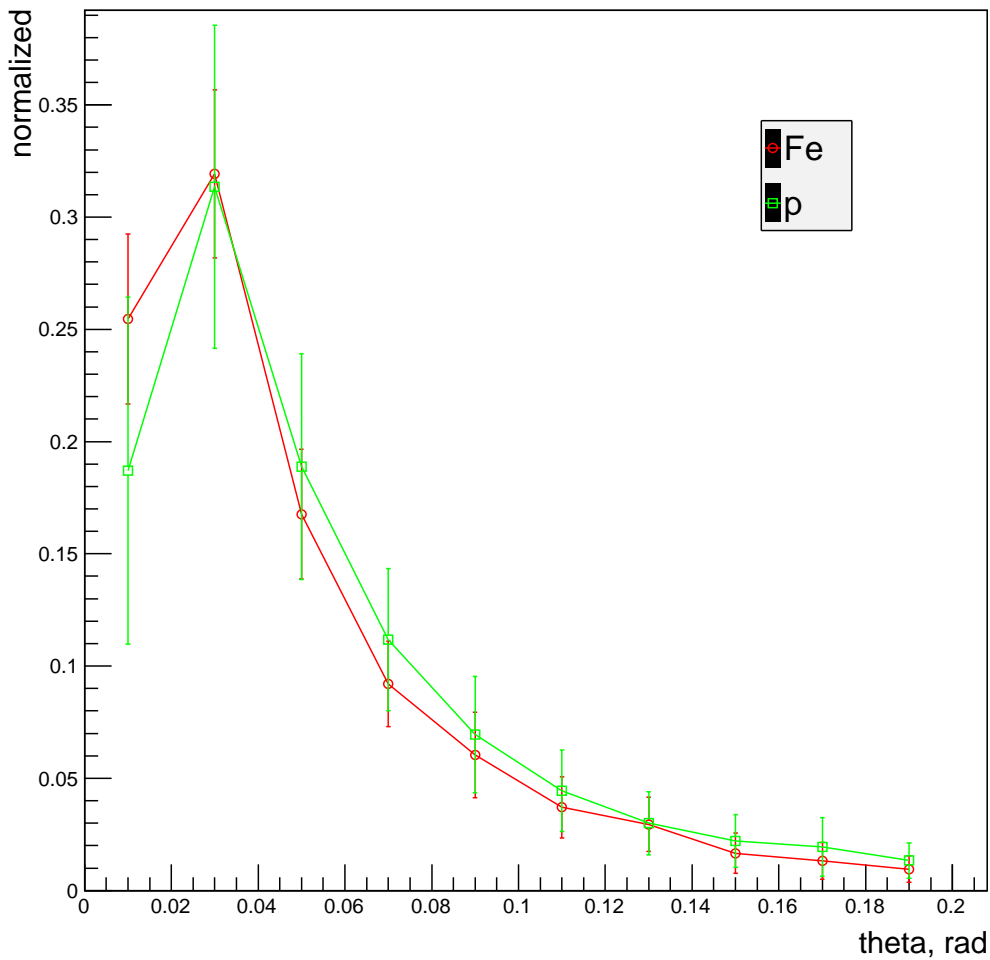}
\caption{\label{l2} Comparison of muon angular distributions for 100 PeV p and Fe initiated shower in case of uniform grid in $\alpha$ (all histogram bins have equal widths). Interaction model: CORSIKA/QGSJET01 .}
\end{figure}

- typical core distance $R$ for the primary energies of interest should ensure, on one hand, large enough number of detected muons (i.e. $R$ being not too large), on the other hand, a sufficient base to scan the muon cascade curve from the side (i.e. $R$ being not too small); \\
- tracker angular resolution should be high enough for the polar angle bins to be made rather narrow to scan at least a few kilometers above the observation level, still the tracker must not be too expensive due to the angular resolution requirements; \\
- energy threshold must ensure that the detected muons can come from high above the observation level being abundant at the same time; \\
- azimuthal angle $\phi$ (see Fig.1) limitations appear due to the fact that the muons of interest should come mainly from the shower core and not from its periphery; \\
- muon energy threshold and azimuthal angle limitations should not reduce the number of detected muons substantially.

Preliminary tests with $E_0$=10 PeV showers showed that the tracker must be of area $S \sim$ 100 m${}^2$ and work at core distance $R \gtrsim 60$ m but the number of detected muons was rather low under such conditions $(E_0,S,R)$ to form angular histograms.

With $E_0=10^{17},10^{18}$ eV we use $S \, =$ 100 m${}^2$, $R \, = \, 100$ m within this work but our approach can definitely be applied to other $S$ and $R$, the above presented considerations taken into account.

Naive muon angular distribution processing includes histogramming all muons of a sample in $\alpha$ above 1 GeV within azimuthal angle $\phi$ band [-0.2,0.2], bins in $\alpha$ being of equal width. An example of $p$ and $Fe$ histogram comparison is shown in Fig.2. Marker abscissa correspond to the centers of histogram bins, marker ordinates represent bin contents averaged over a sample while error bars show r.m.s. deviatons of the contents. One can see no definite separation of the two distributions.

\begin{figure}[h]
\includegraphics[scale=0.85]{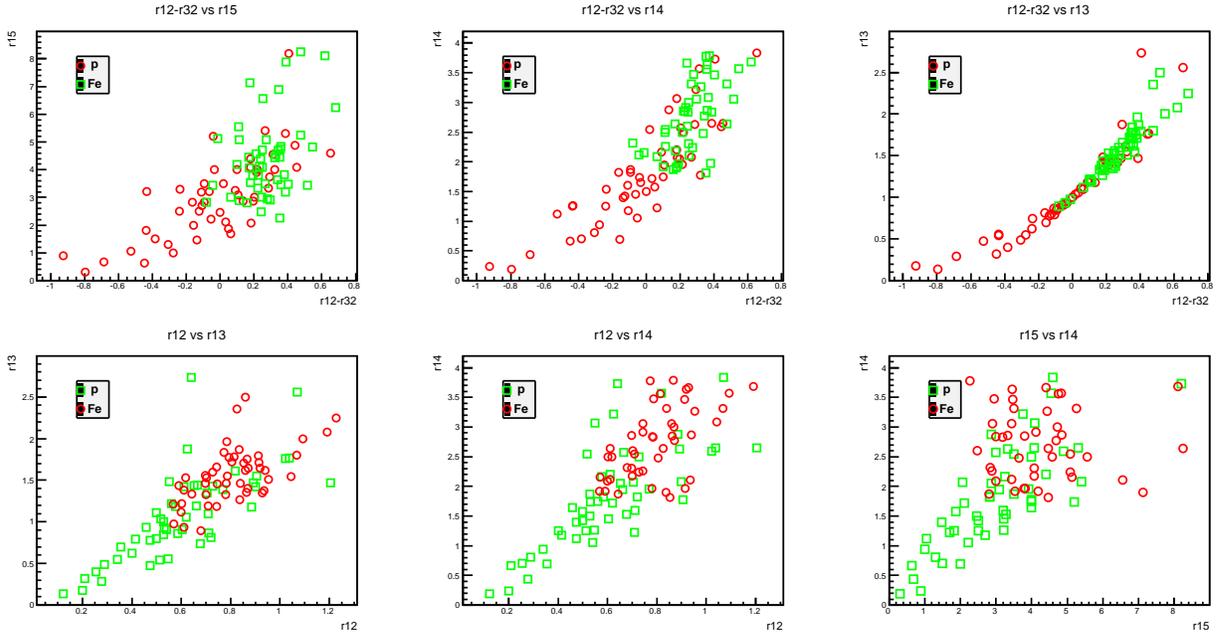}
\caption{\label{l3} Comparison of 100 PeV p and Fe shower samples in different 2-dimensional feature spaces in case of uniform grid in $\alpha$. Features used: $r_{ij}$ and $r_{ij}-r_{kj}$. Interaction model: CORSIKA/QGSJET01 .}
\end{figure}
\begin{figure}[!h]
\includegraphics[scale=0.75]{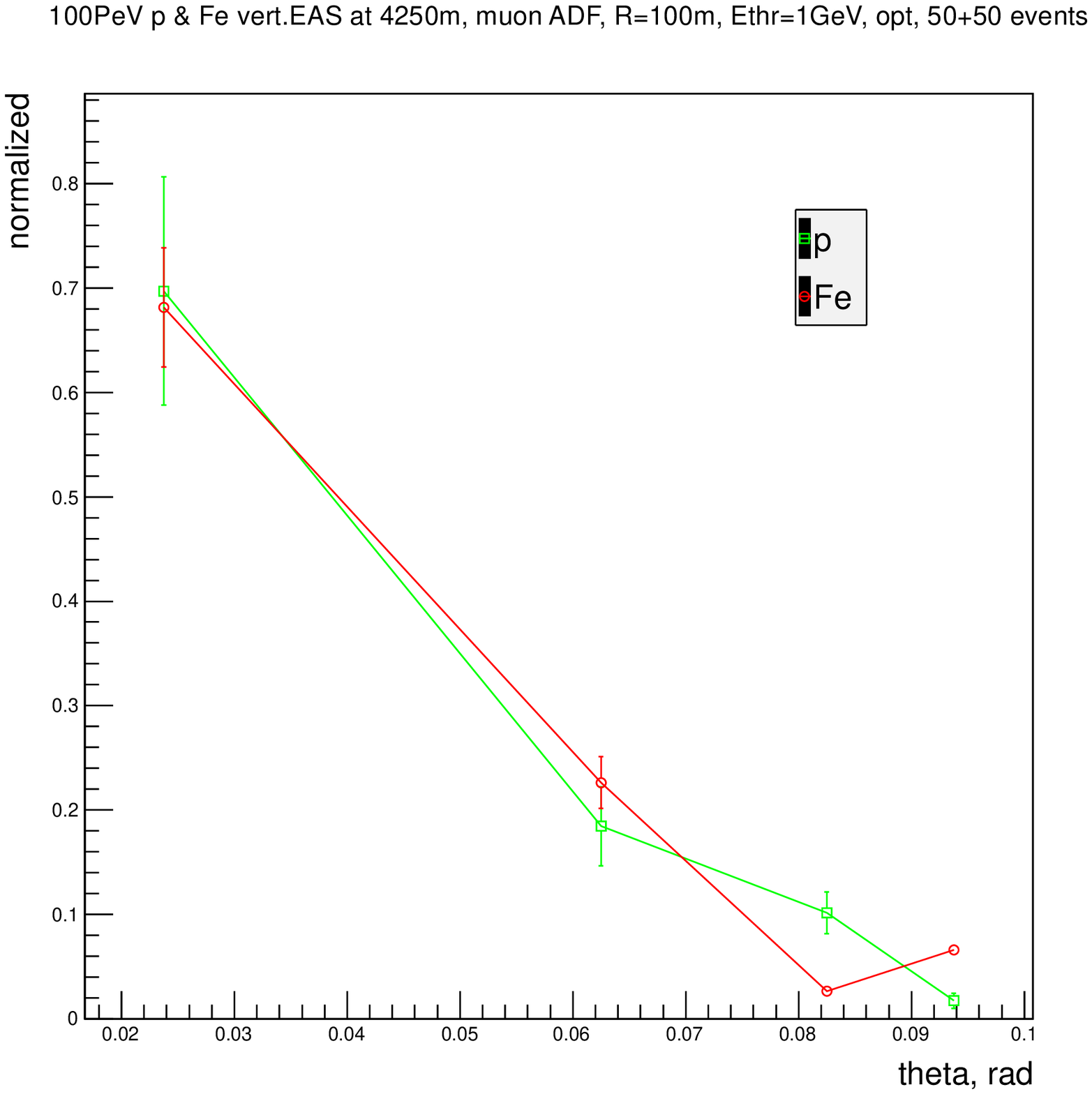}
\caption{\label{l4} Comparison of muon angular distributions for 100 PeV p and Fe initiated shower in case of optimum grid in $\alpha$. Interaction model: CORSIKA/QGSJET01 .}
\end{figure}

Next step of the naive procedure is the presentation of the two samples in 2-dimensional feature space. A few different angular histogram shape measures $r_{ij} \, = \, w_i/w_j$ were tried , here $w_i$ is the i-th bin content of $\alpha-$histogram of a shower. 

\begin{figure}[!h]
\includegraphics[scale=0.85]{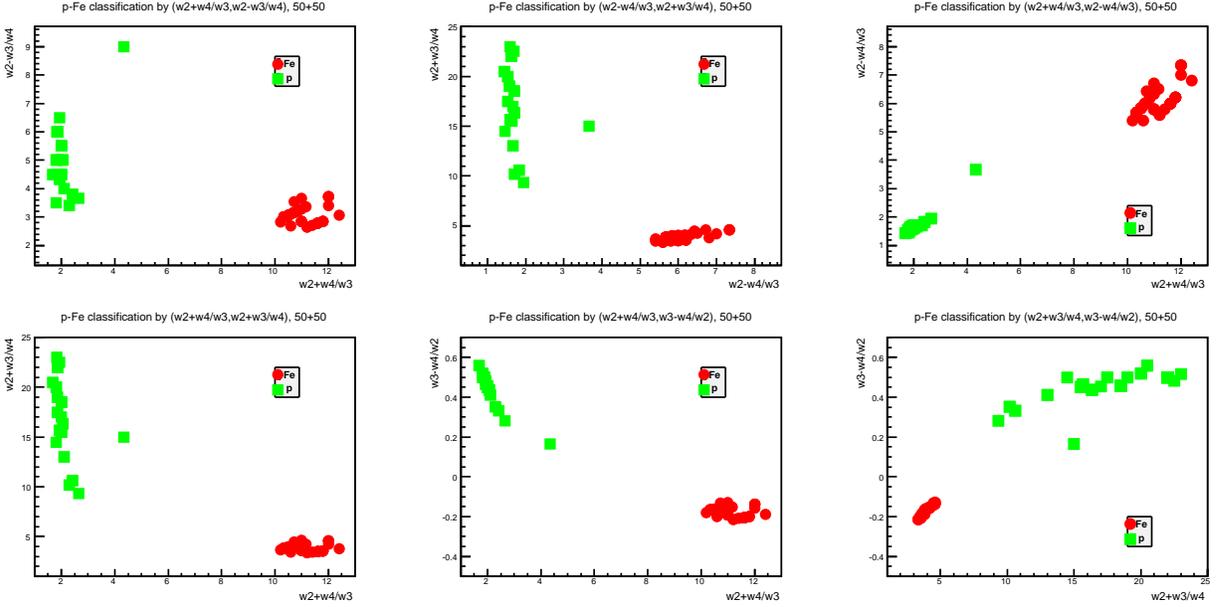}
\caption{\label{l5} Comparison of 100 PeV p and Fe shower samples in different 2-dimensional feature spaces in case of p-Fe-optimum (non-uniform) grid in $\alpha$. Features used: $\rho_{ijk}^{\pm} = \frac{w_i \pm w_j}{w_k}$: $\frac{w_2 + w_4}{w_3} = \rho_{243}^{+}$, $\frac{w_2 - w_4}{w_3} = \rho_{243}^{-}$, etc. Interaction model: CORSIKA/QGSJET01 .}
\end{figure}

\begin{figure}[!h]
\includegraphics[scale=0.85]{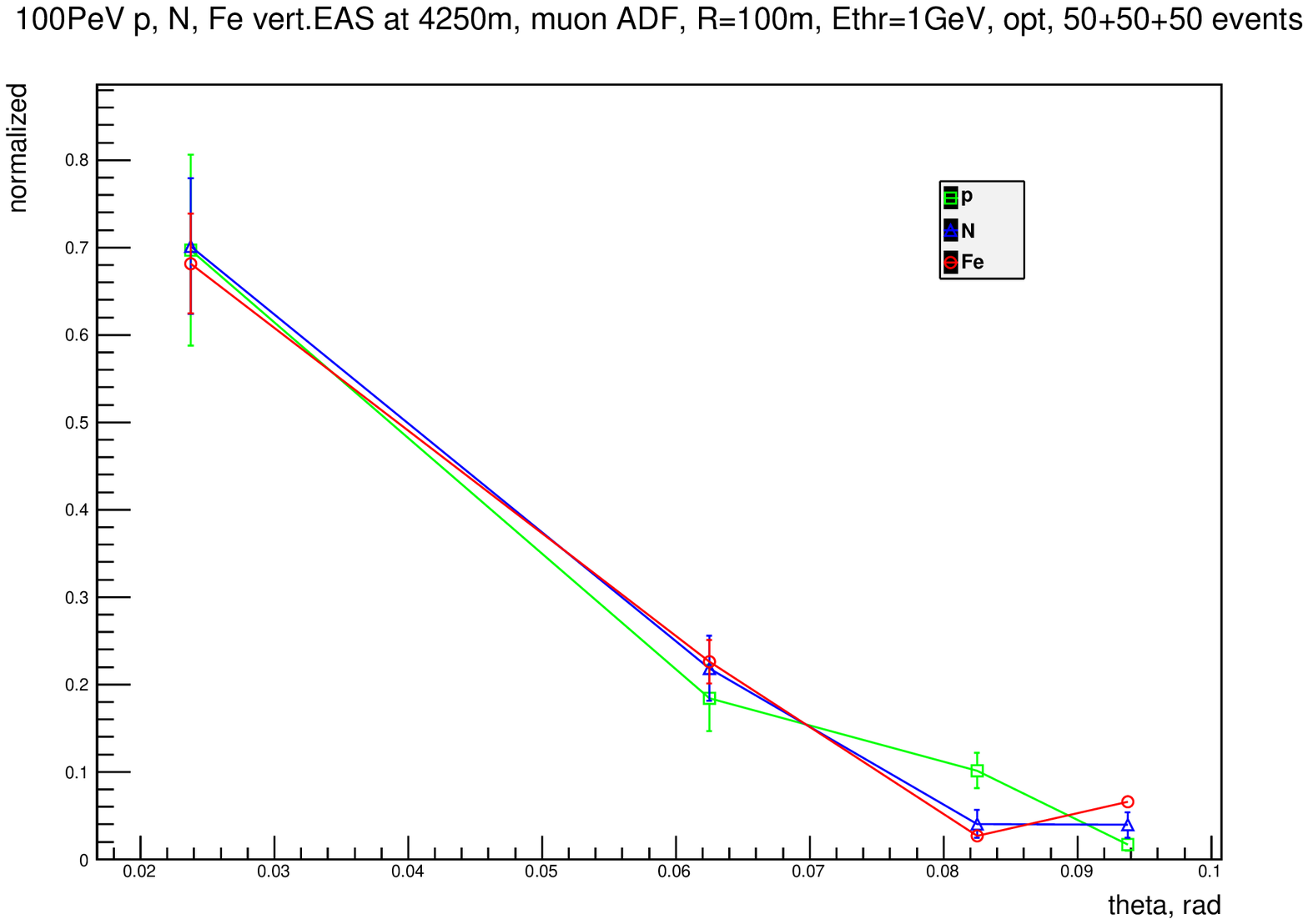}
\caption{\label{l6} Same as Fig.4 but with N-histogram added. Note that the polar angle grid was optimized with respect to p-N separation only.}
\end{figure}

The reason for the use of the distribution shape parameters and not absolute muon bin contents is simple: relative features are less interaction model dependent and are more robust (i.e. less sample dependent). Fig.3 presents some examples of $p$ and $Fe$ samples in 2-dimensional feature spaces: one can see substantial sample overlapping in all spaces so that no obvious criterion for sample separation can be found.

\begin{figure}[!h]
\includegraphics[scale=0.85]{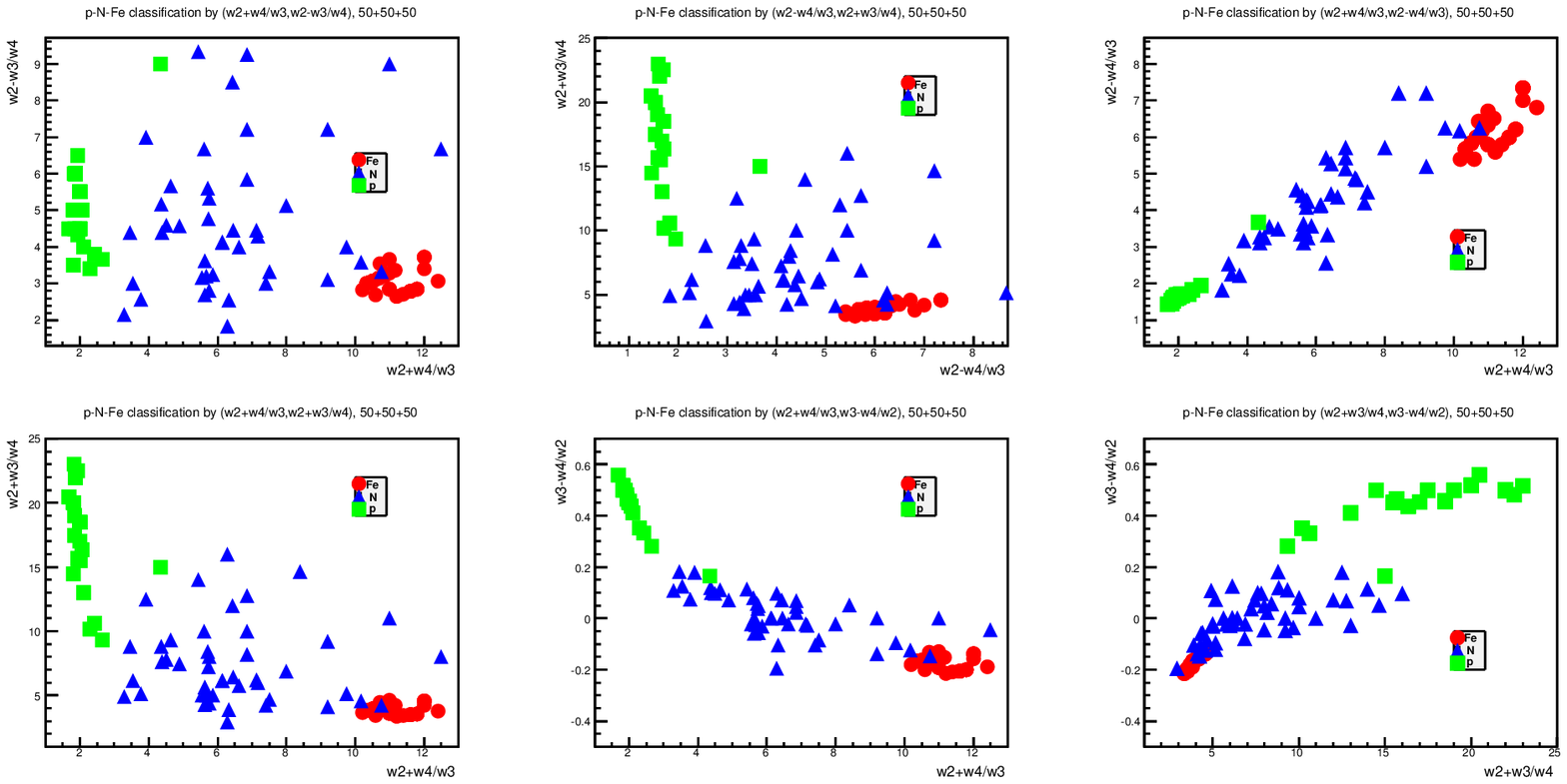}
\caption{\label{l7} Same as Fig.5 but with N-sample added.}
\end{figure}

Discouraging result of the use of the naive procedure forces one to look for some radical improvements of the approach. The first improvement comes from a consideration that equal width bins do not reflect the real scale of the muon cascade and one should optimize the bin grid. The second one takes into account the steep shape of the $\alpha$-distribution and rather large fluctuations of the bin contents: generally speaking, one should reduce the number of histogram bins to 4 or 5 leaving the $\alpha$-range about the same, which will suppress the fluctuations and make the separability of the samples better.

\begin{figure}[!h]
\includegraphics[scale=0.85]{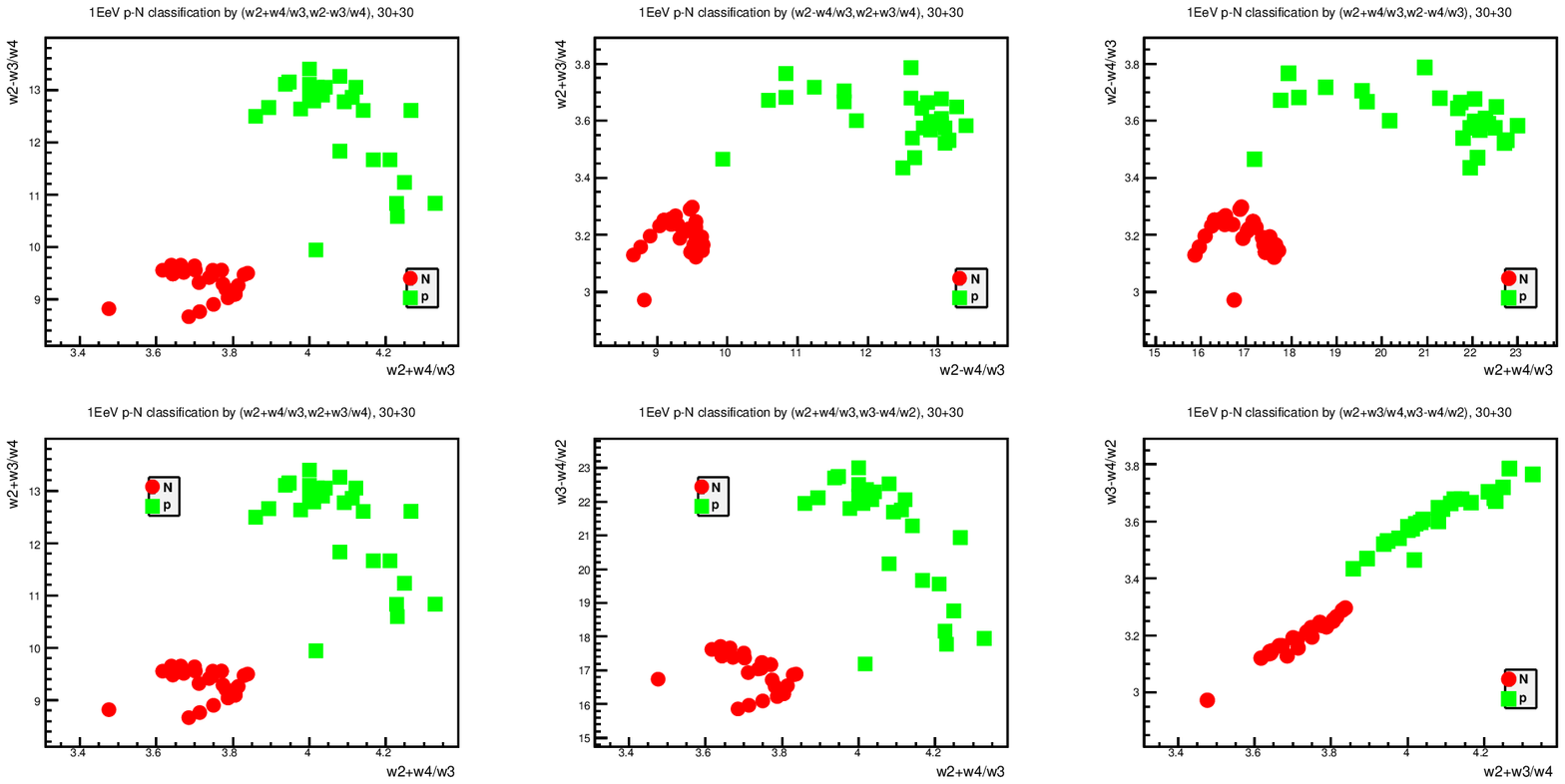}
\caption{\label{l8} Comparison of 1 EeV p and N shower samples in different 2-dimensional feature spaces in case of p-N-optimum grid in $\alpha$. Interaction model: CORSIKA/QGSJET01 .}
\end{figure}

\begin{figure}[!h]
\includegraphics[scale=0.85]{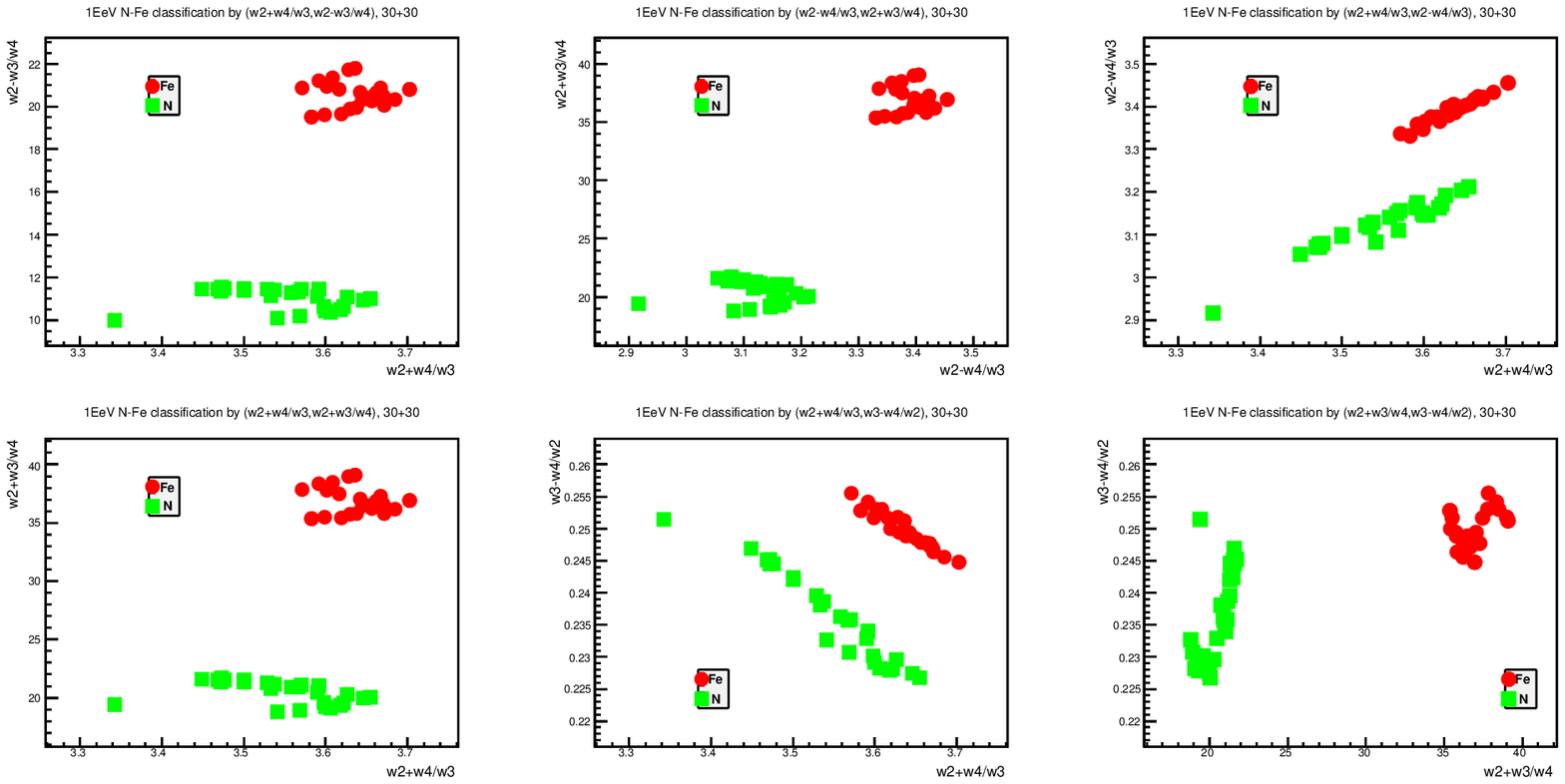}
\caption{\label{l9} Comparison of 1 EeV N and Fe shower samples in different 2-dimensional feature spaces in case of N-Fe-optimum grid in $\alpha$. Interaction model: CORSIKA/QGSJET01 .}
\end{figure}

To carry out the two improvements one is to introduce some quantitative measure of sample separability. We use Bhattacharyya distance $D_B$ under an assumption that the angular bin contents behave as multivariate normal distribution. Each sample is characterized by its own mean vector $\mu_k \equiv  \overline{\vec w_k}$ and covariance matrix $\Sigma_k$, $k=p,Fe$. Then Bhattacharyya distance for p-Fe pair takes the form:
$$
D_B^{p-Fe} \, = \, \frac{1}{8} (\mu_p - \mu_{Fe})^T \Sigma^{-1} (\mu_p - \mu_{Fe}) \, + \, \frac{1}{2} ln \left( \frac{det\Sigma}{\sqrt{det\Sigma_p \cdot det\Sigma_{Fe}}} \right) \, , \;\;\; \Sigma \, = \, \frac{\Sigma_p + \Sigma_{Fe}}{2} \, .
$$
For simplicity one can assume that $\Sigma_k$ matrices are diagonal which means the independence of different bins of the histogram. The latter assumption is not really true but does not change the result substantially.

\begin{figure}[!h]
\includegraphics[scale=0.85]{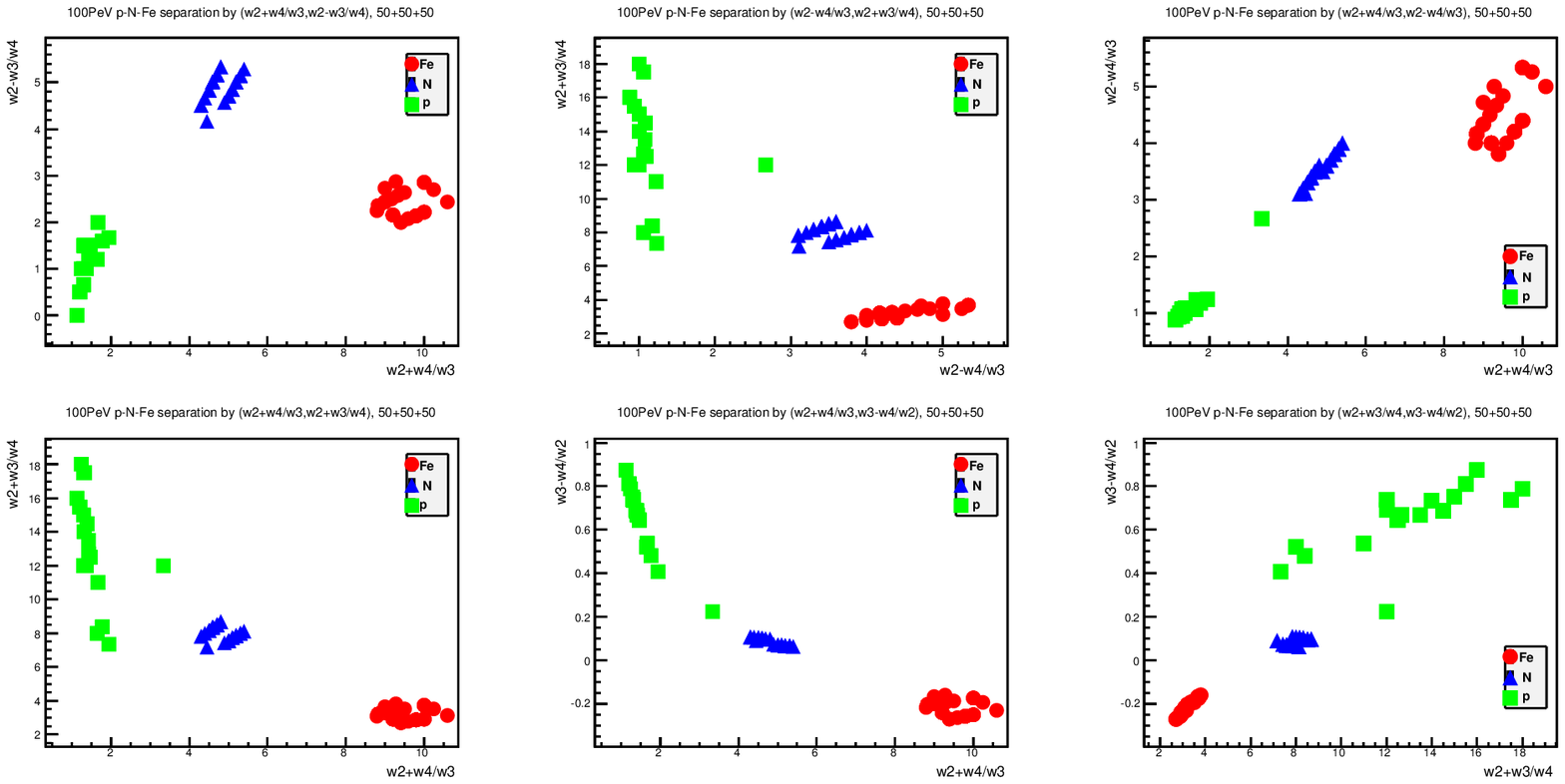}
\caption{\label{l10} Comparison of 100 PeV p, N and Fe shower samples in different 2-dimensional feature spaces in case of $R_{p-N-Fe}$-optimum grid in $\alpha$. Interaction model: CORSIKA/QGSJET01 .}
\end{figure}

After the measure of the class (sample) separability is set one can design a procedure to tune the bin width vector $\vec h$  so that $D_B = D_B \left(\vec h \right)$ approaches its maximum for a concrete pair of samples. Fig.4 shows the result of such optimization for $p-Fe$ pair of $E_0=10^{17}$ eV. One can see definite differences between the two histograms as the bin contents (designated by markers with error bars) do not overlap for bins 2 to 4, unlike the histograms in Fig.2. Thus, we form appropriate histogram shape measures $\rho_{ijk}^{\pm} = \frac{w_i \pm w_j}{w_k}$, $i,j,k=2,3,4$, $i \ne j$, $i \ne k$, $j \ne k$ and present $p$ and $Fe$ samples in different 2-dimensional feature spaces  in Fig.5. Note that the two samples are fully separated for all combinations of $\rho_{ijk}^{\pm}$ considered. Fig.6 shows the same $p$ and $Fe$ histograms as in Fig.4 but with $N$-histogram added. The corresponding feature space presentations are given in Fig.7. Though the bin widths were optimized for better $p-Fe$ separation, $N$-sample overlaps $p$- and $Fe$-samples only partially for all $\rho_{ijk}^{\pm}$ combinations in Fig.7.

\begin{figure}[!h]
\includegraphics[scale=0.85]{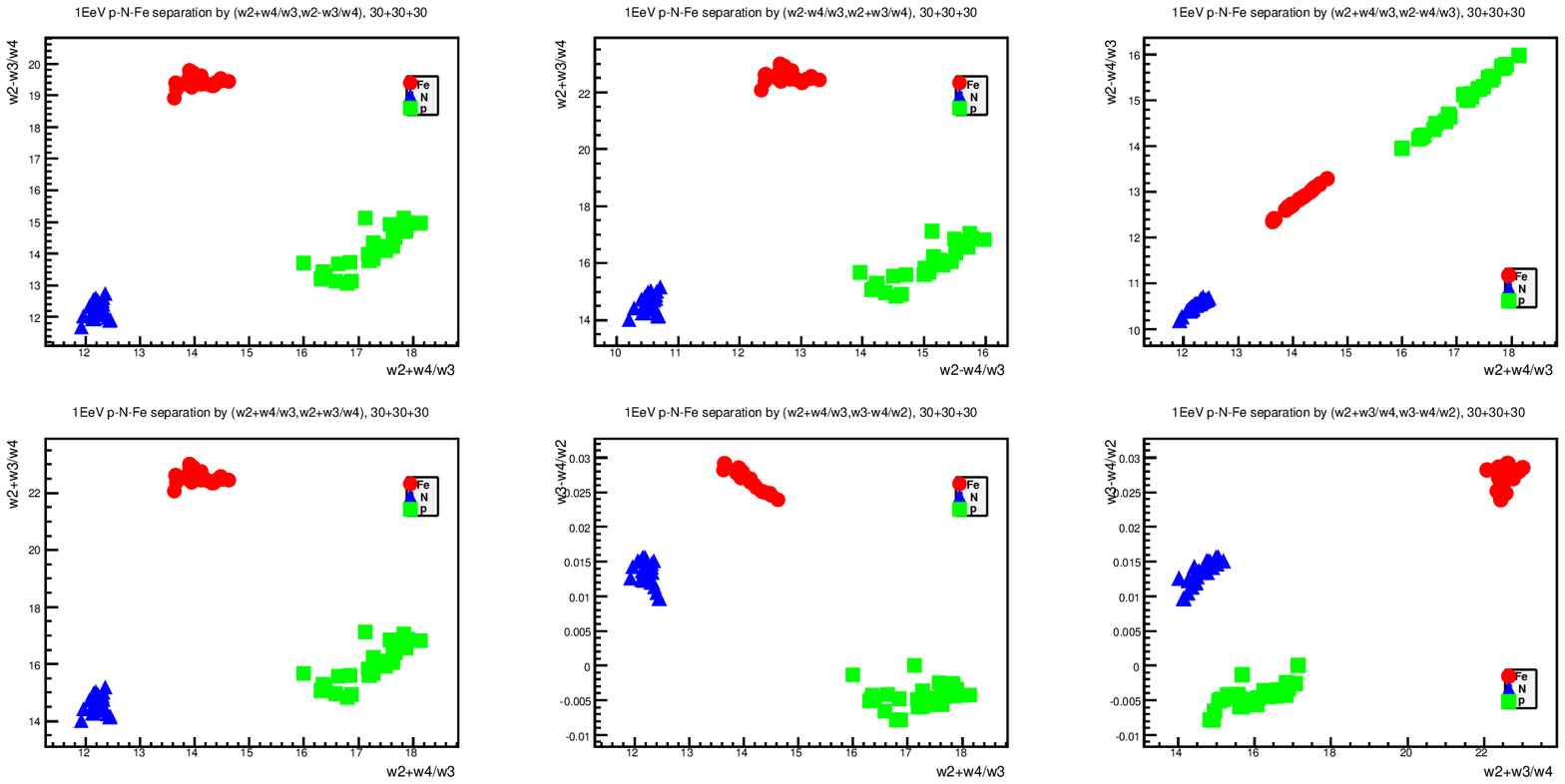}
\caption{\label{l11} Comparison of 1 EeV p, N and Fe shower samples in different 2-dimensional feature spaces in case of $R_{p-N-Fe}$-optimum grid in $\alpha$. Interaction model: CORSIKA/QGSJET01 .}
\end{figure}

One can also optimize the $\alpha$ bin grid for the best separation of $p-N$ or $N-Fe$ sample pairs. The corresponding feature space plots are presented in Fig.8 and Fig.9 for $E_0=10^{18}$ eV.
% For $E_0=10^{18}$ eV $p-Fe$, $p-N$ and $N-Fe$ pairs the 2-dimensional feature space plots are given in Figs 10-12.
Sample separation is good for all $\rho_{ijk}^{\pm}$ combinations presented.
\vspace{0.3cm}

We also carry out one more improvement of the bin width optimization algorithm, namely, a simultaneous maximization of three pair separation while optimizing the widths. Such improvement can be made in many different ways.
We do this as follows: \\

1. calculate $D_{B,max}$ for each of three pairs p-Fe, p-N, N-Fe ($D_{B,max}^{p-Fe}$,
$D_{B,max}^{p-N}$, $D_{B,max}^{N-Fe}$) by varying the vector $\vec h$; \\
2. minimize a function $R_{p-N-Fe}\left( \vec h \right)$
$$
R_{p-N-Fe}\left( \vec h \right) \, = \, \left( \frac{D_B^{p-Fe}}{D_{B,max}^{p-Fe}} - 1 \right)^2 \, + \, \left( \frac{D_B^{p-N}}{D_{B,max}^{p-N}} - 1 \right)^2 \, + \, \left( \frac{D_B^{N-Fe}}{D_{B,max}^{N-Fe}} - 1 \right)^2 \; .
$$

Such algorithm guarantees simultaneous approach of $D_B$ for each pair to its maximum while keeping the contributions of different pairs in balance.
%Fig.10 shows 2-dimensional feature space plots for $E_0=10^{18}$ eV after maximization of $\tilde D_B \, = \, D_B^{p-Fe} \, + \, D_B^{p-N} \, + \, D_B^{N-Fe}$. One can conclude, that for all presented $\rho_{ijk}^{\pm}$ combinations the separation of the three samples is good.
Resulting sample plots for 100 PeV and 1 EeV showers are presented in Fig.10 and Fig.11, respectively. Sample separation is rather good for all 2-dimensional feature spaces considered.
\vspace{0.3cm}

\begin{figure}[!h]
\includegraphics[scale=0.85]{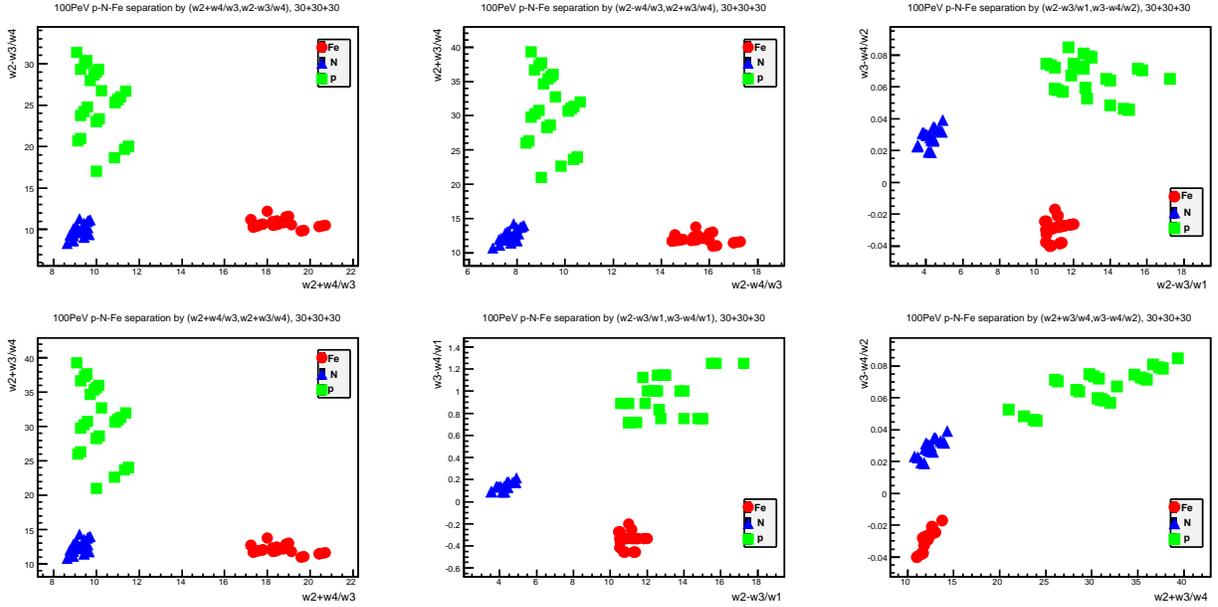}
\caption{\label{l12} Comparison of 100 PeV p, N and Fe shower samples in different 2-dimensional feature spaces in case of $R_{p-N-Fe}$-optimum grid in $\alpha$. Interaction model: CORSIKA/QGSJET-II .}
\end{figure}

Finally, we apply the $R_{p-N-Fe}$-minimization algorithm to 100 PeV p, N and Fe samples made with CORSIKA6.990/QGSJET-II in order to check whether the change of the interaction model can spoil the muon angular distribution sensitivity to the primary particle mass. Fig.12 shows the corresponding sample plots in different 2-dimensional feature spaces. One can see that after the change of the interaction model the sensitivity still persists though the set of the most suitable feature spaces may differ for different models.

\vspace{1.5cm}

\centerline{\bf Conclusion}
\vspace{0.5cm}

1. A new muon tracker approach to PCR mass composition problem in EAS detection is based on high energy ($\gtrsim$ 1 GeV) muon angular distribution measurements by a large area ($\sim$100 m${}^2$) tracking detector of $\sim 0.3 \degree$ angular resolution placed at core distance R$\sim$100 m to view the shower sideways.
\vspace{0.5cm}

2. In order to separate showers initiated by different primary nuclei, a special processing of the muon angular distributions is required which maximizes the differences between event histograms and, thus, makes it possible to distinguish between different groups of primary nuclei.
\vspace{0.5cm}

3. Concrete realizations of event selection algorithm made for "Pamir-XXI" conditions show that "complete" separation of at least three groups of nuclei is possible which makes the new method
% unrivaled
second to none among the charged particle methods in the primary energy range considered.
\vspace{0.5cm}

4. The new approach presumably must work under different observation conditions: observation levels and/or primary energy ranges. Certainly, the optimal $\alpha$-histogram bin widths will differ in different cases.
\vspace{0.5cm}

5. The dependence of the method on the hadron/nucleus interaction model should be studied next but it is clear now that the sensitivity of muon angular distribution to the primary mass will hold because the general properties of the air shower remain the same no matter which interaction model is used.
\vspace{0.5cm}

All in all, large area muon tracker can become an important instrument for the solution of the PCR mass composition problem at $E_0 \gtrsim 10^{17}$ eV.
\vspace{1.5cm}

%\begin{thebibliography}{9}

\centerline{\bf References}
\vspace{0.5cm}
\hspace*{-0.6cm}1.  V.I. Galkin, T.A. Dzhatdoev, 
Moscow Univ. Phys. Bull., Vol. 65, Issue 3, 195-202 (2010). \\
\hspace*{0.5cm}V. I. Galkin, T. A. Dzhatdoev, Bull. Russian Acad. Sci.: Phys., Vol. 75, Issue 3, 309-312 \\
\hspace*{0.7cm}(2011). \\
%\hspace*{0.5cm}A.S. Borisov, V.I. Galkin, J. Phys.: Conf. Ser. 409 012089. \\
2. A.S. Borisov, V.I. Galkin, J. Phys.: Conf. Ser. 409 012089. \\
%\bibitem{KASK} W.D. Apel et al. (KASKADE-Grande Collaboration), Phys. Rev.
3. W.D. Apel et al. (KASKADE-Grande Collaboration), Phys. Rev. Lett. {\bf 107,} 171104 \\
\hspace*{0.7cm}(2011). \\
\hspace*{0.5cm}W.D. Apel et al. (KASKADE-Grande Collaboration), Phys. Rev. D {\bf 87,}  081101(R), (2013).
\hspace*{0.5cm}W.D. Apel et al. (KASKADE-Grande Collaboration), Astropart. Phys. {\bf 47}  (2013) 54-66. \\
4. D.Heck, CORSIKA: A Monte Carlo code to simulate extensive air showers, \\ \hspace*{0.7cm}Report FZKA 6019, Forschungszentrum Karlsruhe, Germany, 1998.
%\end{thebibliography}

\end{document}